# Towards Joint Electricity and Data Trading: A Scalable Cooperative Game Theoretic Approach

Mingyu Yan, *Member, IEEE*, Fei Teng, Senior *Member, IEEE*

*Abstract*—This paper, for the first time, proposes a joint electricity and data trading mechanism based on cooperative game theory. All prosumers first submit the parameters associated with both electricity and data to the market operator. The operator utilizes the public and prosumers' private data to forecast the distributed renewable generators (DRGs) and quantify the improvement driven by prosumers' private data in terms of reduced uncertainty set. Then, the operator maximizes the grand coalition's total payoff considering the uncertain generation of DRGs and imputes the payoff to each prosumer based on their contribution to electricity and data sharing. The mathematical formulation of the grand coalition is developed and converted into a second-order cone programming problem by using an affine-policy based robust approach. The stability of such a grand coalition is mathematically proved, i.e., all prosumers are willing to cooperate. Furthermore, to address the scalability challenge of existing payoff imputation methods in the cooperative game, a two-stage optimization-based approach is proposed, which is converted into a mixed-integer second-order cone programming and solved by the Benders decomposition. Case studies illustrate all prosumers are motivated to trade electricity and data under the joint trading framework and the proposed imputation method significantly enhances the scalability.

## Nomenclature

### 1. Indices and Sets

| | |
|---|---|
| $i, t$ | Indices of prosumers and dispatch periods. |
| $m, w$ | Indices of microturbines (MTs), and distributed renewable generators (DRGs). |
| $C$ | Indices of coalitions. |
| $N$ | **Set of players in the cooperative games.** |

### 2. Parameters

| | |
|---|---|
| $PD_{i,t}^{\min}, PD_{i,t}^{\max}$ | Lower/upper bounds of load for prosumer $i$ at hour $t$. |
| $PW_{i,w,t}^0$ | Power output forecast of DRG $w$ in prosumer $i$ at hour $t$. |
| $PG_{i,m}^{\max}$ | Maximum output of MT $m$ in prosumer $i$. |
| $\lambda_{i,t}, \beta_{i,t}$ | Parameters representing the revenue of MG $i$ at hour $t$. |
| $a_{i,m}, b_{i,m}, c_{i,m}$ | Parameters representing the cost of MT $n$ in MG $i$. |
| $\pi_t^b, \pi_t^s$ | Prices of purchasing and selling energy with the Market Operator (MO) at hour $t$. |
| $\pi_t^{M,up}, \pi_t^{M,dw}$ | Price of purchasing up/down reserve from the MO at hour $t$. |
| $\pi_{i,m,t}^{G,up}, \pi_{i,m,t}^{G,dw}$ | Up/down reserve costs of MT $m$ in prosumer $i$ at hour $t$. |
| $\pi_{i,m,t}^{D,up}, \pi_{i,m,t}^{D,dw}$ | Up/down reserve costs of load in prosumer $i$ at hour $t$. |
| $M$ | Large positive number. |
| $\varepsilon$ | Extremely small positive number. |
| $k^h, k_C$ | Uncertainty budget. |

### 3. Variables

| | |
|---|---|
| $PG_{i,m,t}^0$ | Generation of MT $m$ in prosumer $i$ at hour $t$ in base case. |
| $PD_{i,t}^0$ | Load of prosumer $i$ at hour $t$ in base case. |
| $RM_{i,t}^{up}, RM_{i,t}^{dw}$ | Up/down reserve purchased by prosumer $i$ from the MO at hour $t$. |
| $RD_{i,t}^{up}, RD_{i,t}^{dw}$ | Up/down reserve provided by the load of prosumer $i$ at hour $t$. |
| $RG_{i,m,t}^{up}, RG_{i,m,t}^{dw}$ | Up/down reserve provided by the MT in prosumer $i$ at hour $t$. |
| $\Delta PW_{i,w,t}$ | Power generation deviation from the forecast of DRG $w$ in prosumer $i$ at hour $t$. |
| $RM_{i,t}$ | Reserve provided by the market operator for prosumer $i$ at hour $t$ when power imbalance occurs. |
| $RD_{i,t}$ | Reserve provided by the load of prosumer $i$ at hour $t$ when power imbalance occurs. |
| $RG_{i,m,t}$ | Reserve provided by MT $m$ in prosumer $i$ at hour $t$ when power imbalance occurs. |
| $\Gamma_i^D, \Gamma_{i,m}^G, \Gamma_i^M$ | Adjustment coefficients for the load/MT/MO in the affine robust approach |
| $I_i$ | Binary variable, 1 if prosumer $i$ is in coalition $C$, 0 otherwise. |
| $z_C$ | Binary variable, 1 if prosumers form coalition $C$, 0 otherwise. |
| $IS_{i,t}, IB_{i,t}$ | Auxiliary variables |
| $IG_{i,m,t}, ID_{i,t}$ | Auxiliary variables |

## I. Introduction

CLIMATE change caused by excess carbon emissions has resulted in more intense extreme events and significantly threatened human society over the last decades [1]. Distributed renewable generators (DRGs) can mitigate climate change by providing energy without carbon emission and have attracted growing attention [2]. To accommodate the high penetration of DRGs, transactive energy is proposed to allow geographically-close prosumers flexibly trade energy with their counterparties. Refs.[3]-[4] provided centralized transactive energy markets, in which a central market operator collects bids from all participants and clears the market. Decentralized transactive energy markets were studied in Refs. [5]-[7] to preserve participants' privacy.

However, the DRG's generation is uncertain and hard to be predicted accurately. Prosumers may need to maintain or

M. Yan and F. Teng are with the Department of Electrical and Electronic Engineering, Imperial College London, London SW7 2AZ, U.K. (e-mail: mingyu.yan@imperial.ac.uk, f.teng@imperial.ac.uk).

purchase up/down reserves to mitigate the potential power imbalance caused by the forecast error. Ref. [8] provided a joint transactive energy and reserve trading that adopts the chance-constrained method to accommodate such a uncertainty. Reducing the forecast error for DRGs could significantly reduce the overall cost on reserve and thus raise prosumers' payoffs. In fact, prosumers are geographically close and their onsite DRGs' outputs are correlated. By sharing relevant data (e.g., meteorological parameters or forecast results) with their counterparts, prosumers can reduce the forecast error of their onsite DRGs. Several literatures have already studied how to reduce forecast error by using relevant data from multiple renewable resources. Ref. [9] utilized the geographical data and computational fluid dynamics method to describe the spatial correlation of multiple wind farms. Furthermore, Ref. [10] provided an ellipsoidal uncertain set to describe the potential joint forecast error, which can better reflects the correlation of multiple wind farms than the conventional cubic set. Ref. [11] introduced a data-driven approach to build such ellipsoidal sets.

Although sharing data could significantly reduce operation cost, prosumers do not explicitly gain any rewards by doing so in the existing electricity market framework. Moreover, privacy concerns may prevent prosumers from sharing their data without reward [12]. Therefore, an incentive scheme must be employed to promote the sharing of accurate and relevant data. In contrast to most products, data is zero-marginal cost [13], replicable [14], and redundant [15]. Therefore, the existing market frameworks and pricing mechanisms in the electricity market cannot be directly adopted for data trading. To enable the circulation and transaction of data, the design of data market has started to attract significant attention.

Overall, the relevant studies can be categorized into three types: data collection, data circulation, and data pricing. Data collection focuses on gathering effective data, including raw data collection, noise removal and so on. Data circulation investigates how to ensure that data can be circulated among participants in a secure manner (e.g., protecting data from being counterfeited and duplicated). Data pricing estimates the economic value and determines the price for data, which is the key element to achieve a sustainable data market. Several references have already focused on data pricing which aims at incentivizing participants to share data to improve renewable forecast accuracy. Refs. [16]-[17] studied the auction-based data market, in which participants first estimate the data value and provide corresponding bids, and the central data market operator then clears the market based on the bids. The auction-based data market can be put into two categories: buyer-centric [16] and seller-centric [17]. In the former, the data price is solely determined by the buyer. The buyer provides a bid, and the seller is assumed to be the price-taker. In the latter, the buyer still provides a bid, but the seller is allowed to add the noise to the data based on the offered price and sell the corresponding data to the buyer. Ref. [18] provided a data valuation method for instructing participants to submit the bid in the auction. The Shannon entropy and non-noise ratio are utilized to assess the data quality and a neural network-based engine is then applied to evaluate data. Ref. [19] proposed a clearing method in the auction-based market, where the contribution of each participant's data in enhancing forecast accuracy is quantified and used to price the data along with the bidding price.

However, there are two main issues remaining to be addressed. First, the separated electricity and data trading cannot guarantee joint market equilibrium and social welfare maximization. Currently, the data market is first cleared based on the predicted utility, and the electricity is further cleared based on the data market clearing results. However, electricity and data prices are interdependent. On the one hand, the data price and data sharing could impact the electricity price. For instance, the participants, who spend extra costs to purchase data in the data market, may increase the bidding price in the electricity market and hence impact the electricity price. While at the same time, the system imbalance may be reduced due to data trading and affect the electricity price. On the other hand, the value of data is highly dependent on electricity market conditions. In the time period when the energy is expensive and the reserve is scarce, data prices could increase since reducing forecast error creates more economic value and vice versa. Therefore, the separated trading framework cannot guarantee proper pricing for electricity and data nor achieve the joint market equilibrium. The second issue is that the auction-based data market is inadequate for the prosumer. The conventional electricity prosumer evolves into a joint electricity and data prosumer since it supplies and consumes both electricity and data simultaneously. The prosumer can purchase data from its peers for enhancing the forecast accuracy of its onsite DRGs, while at the same time, the prosumer can sell its data to the other prosumers for earning profits. It becomes an extremely complicated task to create efficient bids under such a setting and the single-auction market cannot effectively incentivize the prosumer to trade data when the prosumer wants to sell and purchase data simultaneously.

Joint electricity and data trading provides a potential solution to address the first issue. In the joint electricity and data trading market, prosumers simultaneously provide the parameters associated with electricity and data to the market operator. The market operator clears the market and determines a single price for electricity and data. The joint trading can achieve the equilibrium of electricity and data markets, thus guaranteeing market stability. As for the second issue, the cooperative game offers a potential solution. Different from the auction, all prosumers cooperate with each other to maximize the total payoff or social welfare. Prosumers do not need to provide bids, and the total payoff is imputed to each participant based on its contribution in creating economic value rather than the bids. More importantly, the cooperative game based market ensures the stability of the market since prosumers would not find alternative pricing which is more economically preferable. Accordingly, some recent references also started to investigate the cooperative game in the data-only market for evaluating the data of electricity consumers [20]-[21], while how to implement such mechanism in joint electricity and data trading is still missing. The studies above also pointed out that the most challenging problem in a cooperative game-based data market is the scalability in the imputation process. This is because the existing imputation methods like Shapley value and Nucleolus require enumeration for all coalitions and have an exponential computation complexity. This may limit the applicability of such a mechanism.

In this context, this paper addresses the above problems by proposing a cooperative game based joint electricity and data

trading and provides a scalable imputation method. The contributions of this paper include:
1) A joint electricity and data trading mechanism based on the cooperative game is proposed, in which all participants are incentivized to form a grand coalition. Prosumers submit the parameters associated with electricity and data to the market operator, who further determines the optimal schedule and allocates the total payoff for each prosumer. This is the first attempt to provide a joint trading mechanism for electricity and data.
2) The mathematical formulation of the grand coalition considering the uncertain generation of DRGs is developed, which is further converted into a second-order cone programming (SOCP) problem using the affine-policy based robust approach. Based on this formulation, the stability of the grand coalition is mathematically proved, i.e., all prosumers are incentivized to cooperate for trading electricity and data.
3) A scalable imputation method is provided, which calculates the imputation by solving a two-stage nonlinear and nonconvex optimization problem rather than enumerating all coalitions. The two-stage problem is reformulated into a mixed-integer second-order cone programming (MISCOP) problem, which is further solved by the Benders decomposition.

The rest of the paper is organized as follows. The joint electricity and data trading based on the cooperative game is formulated in Section II. Section III presents the mathematical model of the grand coalition and proves its stability. Section IV develops the scalable imputation method. Numerical results are presented to illustrate the validity of the proposed method in incentivizing prosumers to trade electricity and data in Section V, followed by the conclusion and further work in Section VI.

## II. Joint Electricity and Data Market

We consider the joint electricity and data trading for multiple prosumers. Each prosumer manages onsite microturbines (MTs), DRGs and the aggregated load. The prosumer evolves into a joint electricity and data prosumer since it can supply and consumer both electricity and data. The two-stage operation method is used for accommodating uncertain generation of DRGs. At the first stage, the power generation of DRGs is forecasted and the uncertainty set is quantified. Then, the MT and load are scheduled based on the forecast to ensure power balance. When the power generation of DRGs deviates from the forecast at the second stage, sufficient up/down reserves will be employed to accommodate the deviation. In this case, enhancing forecasting accuracy could significantly reduce the reserve requirement, thus reducing the overall cost.

### A. Impact of Data Sharing on Uncertainty Set

Here we use robust optimization to determine the reserve requirement to accommodate uncertain generation. In robust optimization, an uncertainty set is required to describe the potential deviation, such as the cubic set shown in Figure 1. The mathematical formulation of the cubic set is provided in (1). Each participant determines the boundary of the cubic set (i.e., $\Delta PW^{max}$) by using its own meteorological data. However, such a cubic set fails to accurately capture the correlated power generation of geographically-close DRGs and thus results in conservative decisions. If prosumers share their relevant data for forecast, more accurate forecast can be employed and new uncertainty set, ellipsoidal set as shown in Figure 1, can thus be obtained [22]. Compared with the cubic set, the ellipsoidal set contains fewer impossible scenarios, and its volume is smaller than that of the cubic set. Mathematical formulation of the ellipsoid set is presented in (2). By enabling prosumers to trade data, the inaccurate cubic set will be replaced by the ellipsoid set, which reduces the conservativeness of the decision and thus decreases the operation cost. However, the introduction of data trading converts the conventional linear constraints of the cubic set into conic constraints of the ellipsoidal set.

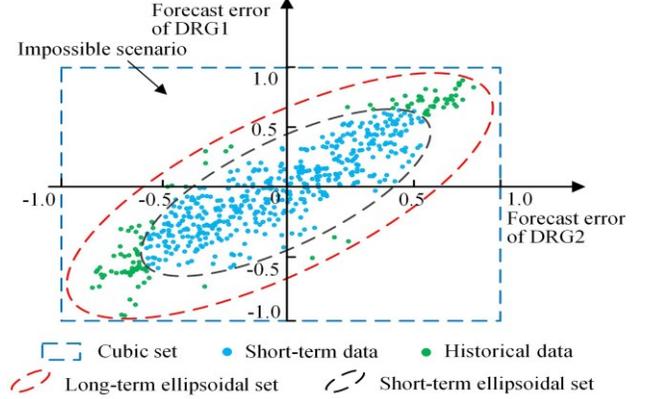

Fig.1 Comparison of box and ellipsoid set

The data markets in the electricity sector can be categorized as long-term and short-term markets. In the long-term data market, prosumers trade historical data (e.g., historical DRG's generation) to determine the ellipsoid set, which is generally used for planning or long-term maintenance. In the short-term data market, prosumers could trade the short-term data including forecast results or relevant meteorological parameters for the next operation hours/days for building the ellipsoid set used in day-ahead/real-time operation. Generally, the volume of the short-term ellipsoid set is much less than that of the long-term ellipsoid set since the short-term data could help eliminate impossible scenarios. Assuming that all prosumers have already shared long-term historical data, we focus on the short-term joint electricity and data market for the day-ahead operation.

$$\Phi := \left\{ \Delta PW \in R^n \middle| \left| \Delta PW \right| \leq \Delta PW^{max} \right\} \quad (1)$$

$$\Phi := \left\{ \Delta PW \in R^n \middle| \left( \Delta PW - c \right)^T Q \left( \Delta PW - c \right) \leq 1 \right\} \quad (2)$$

### B. Joint Electricity and Data Trading

We use the cooperative game to model the joint electricity and data trading market, in which the prosumers are the market participants and are willing to cooperate with their peers. Accordingly, these participants will form a grand coalition. Similar to the data market framework proposed in Ref. [18], there is a centralized market operator in the proposed market that collects relevant parameters (e.g., electricity and data parameters) from each prosumer. The market operator first utilizes the shared data parameters to predict the uncertain generation of DRGs by using various methods, such as the multivariate prediction proposed in [11]. Then, the market operator maximizes the sum of all participants' payoffs and determines the participant's optimal schedule that can accommodate uncertain DRGs based on predicted uncertainty set. Furthermore, the total payoff is allocated to each participant.

In addition to serve as the intermediary that helps electricity and data trading among prosumers, the market operator could trade electricity with the prosumers using the feed-in-tariff mechanism [7] if there is an electricity surplus or deficit.

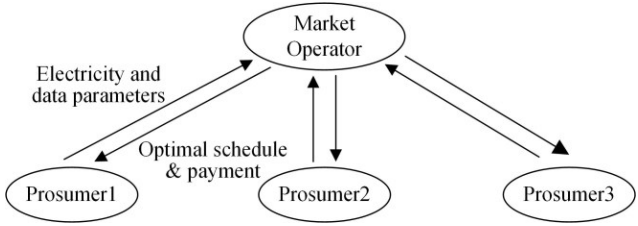

Fig. 2. Structure of the joint electricity and data trading

## III. FORMULATION OF THE GRAND COALITION

### A. Mathematical Formulation

We utilize the two-stage robust optimization to accommodate the uncertain generation of DRGs. During the first stage (i.e., the generation of DRG is equal to the forecast), the total generation and load should be balanced. When generation of DRG is deviated from the forecast, the reserve provided by the market operator, MT and load must be capable to mitigate the power imbalance. The objective of the market operator is to maximize the total payoff, as shown in (3):

$$\max\ U(N) = \sum_i \sum_t \left( R^E(i) - R^{up}(i) - R^{dw}(i) \right) \quad (3)$$

$$R^E(i) = \pi_t^s PS_{i,t} - \pi_t^b PB_{i,t} + R(PD_{i,t}^0) - \sum_m F(PG_{i,m,t}^0) \quad (4)$$

$$R^{up}(i) = \pi_t^{M,up} RM_{i,t}^{up} + \pi_{i,m,t}^{G,up} RG_{i,m,t}^{up} + \pi_{i,t}^{D,up} RD_{i,t}^{up} \quad (5)$$

$$R^{dw}(i) = \pi_t^{M,dw} RM_{i,t}^{dw} + \pi_{i,m,t}^{G,dw} RG_{i,m,t}^{dw} + \pi_{i,t}^{D,dw} RD_{i,t}^{dw} \quad (6)$$

$$U(PD_{i,t}^0) = \lambda_{i,t} PD_{i,t}^0 - \beta_{i,t}(PD_{i,t}^0)^2 \quad (7)$$

$$C(PG_{i,m,t}^0) = a_{i,m}(PG_{i,m,t}^0)^2 + b_{i,m} PG_{i,m,t}^0 + c_{i,m} \quad (8)$$

$$\sum_i (PD_{i,t}^0 + PS_{i,t}) = \sum_i \left( \sum_m PG_{i,m,t}^0 + \sum_w PW_{i,w,t}^0 + PB_{i,t} \right) \forall t \quad (9)$$

$$PD_{i,t}^0 + RD_{i,t}^{up} \leq PD_{i,t}^{\max} \quad \forall i, \forall t \quad (10)$$

$$PD_{i,t}^0 - RD_{i,t}^{dw} \geq PD_{i,t}^{\min} \quad \forall i, \forall t \quad (11)$$

$$PD_{i,t}^0, RD_{i,t}^{up}, RD_{i,t}^{dw} \geq 0 \quad \forall i, \forall t \quad (12)$$

$$PG_{i,m,t}^0 + RG_{i,m,t}^{up} \leq PG_{i,m}^{\max} \quad \forall i, \forall m, \forall t \quad (13)$$

$$PG_{i,m,t}^0 - RG_{i,m,t}^{dw} \geq 0 \quad \forall i, \forall m, \forall t \quad (14)$$

$$PG_{i,m,t}^0, RG_{i,m,t}^{dw}, RG_{i,m,t}^{up} \geq 0 \quad \forall i, \forall m, \forall t \quad (15)$$

$$F := \{ \forall \Delta PW \in \Phi, \exists RG, RD, RM, \\ \text{Constraints } (17)-(21) \} \quad (16)$$

$$\Phi := \left\{ \Delta PW \in R^n \mid (\Delta PW - c)^T Q (\Delta PW - c) \leq 1 \right\} \quad (17)$$

$$\sum_i RD_{i,t} = \sum_i \left( \sum_m RG_{i,m,t} + \sum_w \Delta PW_{i,w,t} + RM_{i,t} \right) \forall t \quad (18)$$

$$-RD_{i,t}^{dw} \leq RD_{i,t} \leq RD_{i,t}^{up} \quad \forall i, \forall t \quad (19)$$

$$-RG_{i,m,t}^{dw} \leq RG_{i,m,t} \leq RG_{i,m,t}^{up} \quad \forall i, \forall m, \forall t \quad (20)$$

$$-RM_{i,m,t}^{dw} \leq RM_{i,m,t} \leq RM_{i,m,t}^{up} \quad \forall i, \forall m, \forall t \quad (21)$$

Objective (3) includes payoff for providing energy, up/down reserve. Please note the data cost is ignored since data is assumed to be a zero-marginal cost product. Constraints (7)-(15) represents the first-stage constraints. Constraint (7) represents the revenue function for balancing flexible load. Constraint (8) represents the quadratic cost function of the MT. Constraint (9) enforces power balance for the grand coalition. Constraints (10)-(15) limit the MT and load while providing reserve. Constraints (16)-(21) limit operation when the deviation of DRG's generation occurs. Constraint (17) indicates the ellipsoidal uncertain set for describing forecast error with data trading. Constraint (18) enforces power balance when DRG's generation deviation occurs. Constraints (19)-(21) ensure that the re-dispatch value provided by MT, load, and market operator must be less than the reserve.

### B. SOCP Reformulation Using Affine Robust Method

The proposed grand coalition model cannot be solved directly due to the consideration of uncertain variable $\Delta PW$. Therefore, the affine robust method (also known as linear decision rule) is adopted to reformulate the primal problem into a computational tractable problem. More specifically, the affine robust method allocates the deviated DRG's generation to the reserves of MTs, loads and the market operator by introducing adjustment coefficients $\Gamma_i^D$, $\Gamma_{i,m}^G$, and $\Gamma_i^M$. Accordingly, (18)-(21) are reformulated as follow:

$$\sum_i \Gamma_i^D + \sum_i \Gamma_i^U + \sum_m \sum_i \Gamma_{i,m}^G = 1 \ \forall i, \forall t \quad (22)$$

$$0 \leq \Gamma_i^U, \Gamma_i^D, \Gamma_i^U \leq 1 \ \forall i \quad (23)$$

$$\Gamma_i^D \sum_i \sum_w \Delta PW_{i,w,t} \leq RD_{i,t}^{up} \quad \forall i, \forall t \\ -\Gamma_i^D \sum_i \sum_w \Delta PW_{i,w,t} \leq RD_{i,t}^{dw} \quad \forall i, \forall t \quad (24)$$

$$\Gamma_{i,m}^G \sum_i \sum_w \Delta PW_{i,w,t} \leq RG_{i,m,t}^{up} \quad \forall i, \forall m, \forall t \\ -\Gamma_{i,m}^G \sum_i \sum_w \Delta PW_{i,w,t} \leq RG_{i,m,t}^{dw} \quad \forall i, \forall m, \forall t \quad (25)$$

$$\Gamma_i^M \sum_i \sum_w \Delta PW_{i,w,t} \leq RM_{i,t}^{up} \quad \forall i, \forall t \\ -\Gamma_i^M \sum_i \sum_w \Delta PW_{i,w,t} \leq RM_{i,t}^{dw} \quad \forall i, \forall t \quad (26)$$

Constraints (22)-(23) enforce that the deviation is exactly allocated to each MT, load and the market operator. Constraints (24)-(26) represents the redispatch value after utilizing the affine robust approach. Accordingly, the primal problem is reformulated as follow:

$$\max\ (3) \\ \text{Constraints } (4)-(15), (22)-(23) \\ F := \{ \forall \Delta PW \in \Phi : \text{Constraints } (17), (24)-(26) \} \quad (27)$$

(27) is a nonlinear and nonconvex problem owing to the uncertain part (i.e., nonlinear constraints (17) and (24)-(26)). First, we present the compact form of constraints (17) and (24)-(26), which is shown as follow:

$$F := \{\forall \Delta PW \in \Phi : (\Delta PW - c)^T Q (\Delta PW - c) \leq 1 \quad A \cdot \Gamma \cdot B \cdot \Delta PW \leq D \} \quad (28)$$

where $A$, $B$, and $D$ are constant matrix. (28) can be reformulated as (29) [23]:

$$(\Delta PW - c)^T Q (\Delta PW - c) \leq 1$$
$$\max(A \cdot \Gamma \cdot B \cdot \Delta PW) \leq D \quad (29)$$

Since (17) is the ellipse, $Q$ is the symmetric positive matrix and can be decomposed into a triangular matrix and its conjugate component (i.e., $L^T L$) using the Cholesky decomposition. Accordingly, constraint (17) can be reformulated as follow:

$$(\Delta PW - c)^T Q (\Delta PW - c) \leq 1$$
$$\Rightarrow (\Delta PW - c)^T (L^T L)(\Delta PW - c) \leq 1 \quad (30)$$
$$\Rightarrow (L(\Delta PW - c))^T (L(\Delta PW - c)) \leq 1$$

By introducing the auxiliary variable $\xi$ defined in (31), (29) can be rewritten as (32)-(33):

$$\xi = L(\Delta PW - c) \quad (31)$$
$$\xi^T \xi \leq 1 \quad (32)$$
$$\max(A \cdot \Gamma \cdot B \cdot (L^{-1}\xi + c)) \leq D \quad (33)$$

(32)-(33) can be further reformulated as (34)-(35), which can then be transformed into conic constraint (36) [24]. Accordingly, the primal nonconvex problem consisting of (3)-(21) can be reformulated as the SOCP problem shown as (37).

$$\|\xi\|_2 \leq 1 \quad (34)$$
$$\max(A \cdot \Gamma \cdot B \cdot L^{-1} \cdot \xi) \leq D - A \cdot \Gamma \cdot B \cdot c \quad (35)$$
$$\|A \cdot \Gamma \cdot B \cdot L^{-1}\|_2 \leq D - A \cdot \Gamma \cdot B \cdot c \quad (36)$$

$$\max (3)$$
$$\text{s.t. } (4)-(15),(22)-(23),(36) \quad (37)$$

(37) is a SCOP problem that can be solved by the existing commercial solver. To improve computational performance, quadratic functions (7)-(8) and conic constraint (36) can be linearized using the methods provided in [7] and [25].

### C. Grand Coalition Stability

For simplicity of the analysis, the linear compact form of the grand coalition operation problem (37) is first shown as (38). The whole problem (38) is defined as $P(N)$. Several definitions and lemmas are provided before proving the stability of the grand coalition.

$$\max U(N)$$
$$\text{s.t. } J(N) \quad (38)$$

**Definition 1 (Imputation):** $x$ represents the payment to each prosumer. The payment can be defined as an *imputation* if following criteria are met:

1) (*Efficiency*) The efficiency of payment $x$ is guaranteed when total payoff is exactly assigned to each participant, i.e., $\sum_i x_i = U(N)$.

2) (*Individual Rationality*) The payment $x$ guarantees the individual rationality if each participant gains more economic benefits after joining the grand coalition, i.e., $x_i \geq U(\{i\}) \ \forall i$.

**Definition 2 (Superadditivity):** The cooperative game is superadditive if (39) is satisfied for any coalitions $C_1, C_2 \subseteq N$:

$$P(C_1 \cup C_2) \geq P(C_1) + P(C_2) \quad (39)$$

**Definition 3 (Positive homogeneity):** If (40) is satisfied for any coalitions $C \subseteq N$ and any constant $\alpha$, :

$$P(\alpha C) = \alpha P(C) \quad (40)$$

**Definition 4 (Core):** *Core* is defined as a set. Imputation belongs to the *Core* only if following equation is satisfied:

$$\left\{ x \mid \sum_i x_i = U(N), \sum_{i \in C} x_i \geq U(C) \ \forall C \right\} \quad (41)$$

The definition of *core* indicates that market participants can always create more economic values by cooperation. Accordingly, all participants will join the grand coalition and accept the imputations in *core*.

**Definition 5 (Balanced game):** The balanced game is a cooperative game that satisfies (42) for any balanced map $\alpha$:

$$\sum_C \alpha_C P(C) \leq P(N) \quad (42)$$

where $\alpha$ is enforced by the following equation:

$$\sum_C \alpha_C I_{i,C} = 1 \quad (43)$$

$$I_{i,C} = \begin{cases} 0 & \text{if } i \notin C \\ 1 & \text{if } i \in C \end{cases} \quad (44)$$

**Definition 6 (Excess):** The *excess* for coalition $C$ is defined as (45):

$$\varepsilon(x, C) = U(C) - \sum_{i \in C} x_i \quad (45)$$

The *excess* is used to measure the dissatisfaction of coalition to the imputation. The positive value of *excess* for coalition $C$ indicates that the members in coalition $C$ are dissatisfied with the existing imputation and would not join the grand coalition. According to the definition of *core*, imputations in the *core* guarantees that the *excess* for each coalition is negative.

**Lemma 1:** The *core* of the balanced game is a non-empty set [26].

**Lemma 2:** If the *core* is a non-empty set, all participants are willing to cooperate, i.e., the grand coalition is stable [27].

**Theorem 1:** The cooperative game for the joint electricity and data trading is superadditive and positive homogeneous.

**Proof:** First, we prove the positive homogeneity of the proposed cooperative game. Since problem (38) is linear programming, equation (46) is satisfied. Accordingly, criterion (40) is satisfied and the proposed cooperative game is positive homogeneous.

$$\alpha U(C) = U(\alpha C), \alpha J(C) = J(\alpha C) \quad (46)$$

Then, we prove the superadditivity of the cooperative game. We first analyze the objective of $P(C_1 \cup C_2)$. $R^E$ for $P(C_1 \cup C_2)$ is formulated as follow:

$$R^E(C_1 \cup C_2)$$
$$= \sum_{i \in (C_1 \cup C_2)} \sum_t \left( \pi_t^s PS_{i,t} - \pi_t^b PB_{i,t} + U_{i,t}(PD_{i,t}^0) - \sum_m C(PG_{i,m,t}^0) \right)$$
$$= \sum_{i \in C_1} \sum_t \left( \pi_t^s PS_{i,t} - \pi_t^b PB_{i,t} + U_{i,t}(PD_{i,t}^0) - \sum_m C(PG_{i,m,t}^0) \right)$$
$$+ \sum_{i \in C_2} \sum_t \left( \pi_t^s PS_{i,t} - \pi_t^b PB_{i,t} + U_{i,t}(PD_{i,t}^0) - \sum_m C(PG_{i,m,t}^0) \right)$$
$$= R^E(C_1) + R^E(C_2) \quad (47)$$

We can obtain similar results for $R^{up}$ and $R^{dw}$, respectively. Accordingly, the following equation is obtained.
$$U(C_1 \cup C_2)$$
$$= R^E(C_1 \cup C_2) - R^{up}(C_1 \cup C_2) - R^{dw}(C_1 \cup C_2)$$
$$= R^E(C_1) - R^E(C_2) - R^{up}(C_1) \quad (48)$$
$$+ R^{up}(C_2) - R^{dw}(C_1) - R^{dw}(C_2)$$
$$= U(C_1) + U(C_2)$$

It can be observed from (48) that the mathematical forms of $U(C_1 \cup C_2)$ and $U(C_1)+U(C_2)$ are exactly the same. Such a conclusion can be extended to most constraints like (9)-(15). Therefore, most constraints of $P(C_1 \cup C_2)$ and $P(C_1)+P(C_2)$ are the same except constraints (9) and (17). For simplicity, we only analyze constraint (9) and constraint (17) can be analyzed similarly. Constraint (9) for $P(C_1 \cup C_2)$ and $P(C_1)+P(C_2)$ can be reformulated as (49)-(50):

$$\sum_{i \in (C_1 \cup C_2)} \left( PD_{i,t}^0 + PS_{i,t} - \sum_m PG_{i,m,t}^0 + \sum_w PW_{i,w,t}^0 + PB_{i,t} \right) = 0 \quad (49)$$

$$\sum_{i \in C_1} \left( PD_{i,t}^0 + PS_{i,t} - \sum_m PG_{i,m,t}^0 + \sum_w PW_{i,w,t}^0 + PB_{i,t} \right) = 0$$
$$\sum_{i \in C_2} \left( PD_{i,t}^0 + PS_{i,t} - \sum_m PG_{i,m,t}^0 + \sum_w PW_{i,w,t}^0 + PB_{i,t} \right) = 0 \quad (50)$$

It is obvious that the solution obtained by (50) is a subset of (49). In other words, the solution space is expanded when constraint (50) is replaced by (49). Therefore, $P(C_1 \cup C_2)$ has a larger solution space than that of $P(C_1)+P(C_2)$ since they have the same objective and most constraints except that $P(C_1 \cup C_2)$ has (49). A larger solution space introduced by (49) indicates a potential higher objective and thus satisfies (39). Accordingly, the superadditivity is proved.

**Theorem 2:** The grand coalition for joint electricity and data trading is stable.

**Proof:** The left side of (42) can be derived as follow based on *Theorem 1*.
$$\sum_{C \in 2^N} \beta(C) P(C) = \sum_{C \in 2^N} P(\beta(C) \cdot C) \text{ Positive homogeneity}$$
$$\leq P\left( \sum_{C \in 2^N} \beta(C) \cdot C \right) \text{ Superadditivity}$$
$$= P\left( \sum_{C \in 2^N} \beta(C) \cdot \sum_i I_i^C \{i\} \right) = P\left( \sum_i \sum_{C \in 2^N} \alpha(C) I_i^C \{i\} \right) \quad (51)$$
$$= P\left( \sum_i \{i\} \right) = P(N)$$

Finally, we prove that the left hand of (42) is smaller than $P(N)$. Therefore, the proposed cooperative game is balanced and, according to *Lemmas* 1 and 2, has the *core*, indicating the stability of the grand coalition.

## IV. SCALABLE IMPUTATION METHOD

### A. Two-Stage Imputation Problem

The market operator fairly allocates the total payoff to each prosumer after obtaining the total payoff in Section III. Sharply value and Aumann-Shapley are the most popular imputation methods in the cooperative game. However, these methods cannot guarantee that the imputation belongs to the *core* and thus may result in an unstable grand coalition. Nucleolus addresses the drawbacks of Sharply value by calculating the imputations from the *core*. However, nucleolus needs to enumerate all potential coalitions and calculating their payoffs. Such an enumeration process requires solving at least $2^N-1$ optimization problem and is thus computational intractability in practice. We provide the following two-stage imputation method that avoids enumeration while ensuring that the imputation belongs to the *core*. The total payoff is imputed to each participant at the first stage, while the second stage checks whether the imputation obtained at the first stage belongs to the *core*. The mathematical model of this two-stage imputation problem is:

$$\min_x \max_I \left( U(C) - \sum_{i \in C} I_i x_i \right) \quad (52)$$

$$\sum_i x_i = U(N) \quad (53)$$

$$U(C) - \sum_i I_i x_i \leq 0 \quad (54)$$

$$1 \leq \sum_i I_i \leq N-1 \quad (55)$$

$$U(C) = \sum_i \sum_t I_i \left( R^E(i) - R^{up}(i) - R^{dw}(i) \right) \quad (56)$$

$$\sum_i I_i \left( \sum_m PG_{i,m,t}^0 + \sum_w PW_{i,w,t}^0 + PB_{i,t} - PD_{i,t}^0 - PS_{i,t} \right) = 0 \; \forall t \quad (57)$$

$$\sum_i I_i \left( \Gamma_i^D + \Gamma_i^U + \sum_m \Gamma_{i,m}^G \right) = 1 \; \forall i, \forall t \quad (58)$$

Constraints $(4)-(15),(22)-(23)$ \quad (59)

$$F := \{ \forall \Delta PW \in \Phi, \exists RG, RD, RU, \text{Constraints } (18)-(21) \} \quad (60)$$

$$\Phi = \left\{ \Delta PW \in R^n \mid (\Delta PW - c)^T Q (\Delta PW - c) \leq k^h - \sum_C k_C z_C \right\} \quad (61)$$

$$z_C = \prod_{i \in C} (1 - I_i) \; \forall C \quad (62)$$

The objective (52) is the same as the objective used in nucleolus, i.e., minimizing the maximum excess among all coalitions. Constraints (53)-(54) enforce that the obtained imputation belongs to the *core*. Constraint (55) restricts the coalition's members. The payoff of coalition $C$ is calculated by using constraint (56). Constraints (57)-(59) represent operation limits for coalition *C* during the base case. Constraint (57) enforces the power balance. Constraint (58) enforces that the participants inside the coalition can share the reserve to mitigate the deviation of DRG's generation. Constraint (60) represents operation limits for coalition *C* when DRG's power generation deviates from the predicted value. Constraints (61)-(62) illustrate the uncertainty set under data trading. We consider that prosumers have already shared historical data and build the uncertainty set shown in (61). $k^h$ represents the uncertainty budget using the historical data. Prosumers further trade data in short-term data market to further reduce the uncertainty budget. $k_C$ is a positive constant that represents the contribution in the data provided by coalition *C* to reduce the uncertainty budget. It can be observed that the uncertain budget (i.e., the right side of (61)) will decrease if prosumers trade their data in the short-term market. $k_C$ can be easily calculated using the method provided in [19] after the market operator collects the data from all prosumers. Constraint (62) enforces the members in coalition *C*.

Constraints (56)-(58) include bilinear product of binary and continuous variables, which can be linearized using the big-M method. For ease of analysis, we only present how to linearize constraint (56) and the other constraints can be linearized similarly. By introducing auxiliary variables $IS_{i,t}$, $IB_{i,t}$, $IG_{i,m,t}$, and $ID_{i,t}$, the bilinear terms $I_i PS_{i,t}$, $I_i PB_{i,t}$, $I_i PG_{i,m,t}$, and $I_i PD_{i,t}$ can be linearized as (63)-(66). By replacing the bilinear terms by auxiliary variables, constraint (56) can be reformulated as a linear constraint. Other constraints in (60) to accommodate uncertain DRG's generation can be reformulated into conic constraints using the affine robust approach in Section III-B.

$$-MI_i \leq IS_{i,t} \leq Mz_i \\ -M(1-I_i) \leq IS_{i,t} - PS_{i,t} \leq M(1-I_i) \quad (63)$$

$$-MI_i \leq IB_{i,t} \leq MI_i \\ -M(1-I_i) \leq IB_{i,t} - PB_{i,t} \leq M(1-I_i) \quad (64)$$

$$-MI_i \leq IG_{i,m,t} \leq MI_i \\ -M(1-I_i) \leq IG_{i,m,t} - PG_{i,m,t} \leq M(1-I_i) \quad (65)$$

$$-MI_i \leq ID_{i,t} \leq MI_i \\ -M(1-I_i) \leq ID_{i,t} - PD_{i,t} \leq M(1-I_i) \quad (66)$$

### B. Scalable Solution Techniques

The primal two-stage imputation problem can be reformulated as a MISCOP problem, which can be rewritten into a compact form (67).

$$\min_{x} \max_{y,z} (A \cdot x + B \cdot y + C \cdot z) \\ s.t. \ D \cdot x \leq H, \ E \cdot y + \|F \cdot z\|_2 \leq G \quad (67)$$

where continuous variable *x* is a set of variables determined at the first stage. Binary variable *y* and continuous variable *z* are the sets of variables determined at the second stage. $D \cdot x \leq H$ represents first-stage constraint (53). $D \cdot y + \|E \cdot z\|_2 \leq G$ represents second stage constraints (54)-(62). We utilize the Benders decomposition to solve the proposed two-stage imputation method, which decouples the original two-stage problem into two single-stage problems, i.e., the master problem and subproblem. The flowchart is shown in Figure 3.

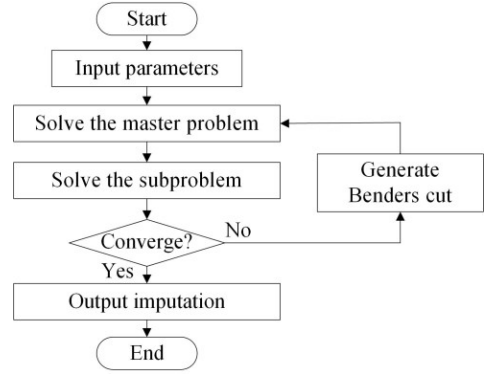

Fig. 3. Flowchart of the Benders decomposition

### C. Master Problem

The grand coalition's payoff is assigned to each participant in the master problem, the mathematical formulation of which is presented as follow:

$$\min_{x} \mu \\ s.t. \ D \cdot x \leq H, \ \mu \geq \mu^{\min} \quad (68)$$

### D. Subproblem

The subproblem aims at finding the coalition with the largest excess under the calculated imputation from (68). The mathematical formulation is presented as follow:

$$\max_{y,z} (A \cdot x^* + B \cdot y + C \cdot z) \\ s.t. \ E \cdot y + \|F \cdot z\|_2 \leq G \quad (69)$$

The optimal solutions $y^*$ and $z^*$ are obtained after solving (69). If criteria (70) is satisfied, the optimal imputation is obtained and the algorithm is terminated. Otherwise, add Benders cut (71) to the master problem:

$$\left| \alpha^* - (A \cdot x^* + B \cdot y^* + C \cdot z^*) \right| \leq \varepsilon \quad (70)$$

$$\alpha \geq A \cdot x^* + B \cdot y^* + C \cdot z^* \quad (71)$$

where $\varepsilon$ is an extremely small number.

### E. Solution Procedure

The proposed solution procedures based on the Benders decomposition are listed as follow:
1. Input all parameters.
2. Solve master problem (68) and obtain the optimal solution.
3. Solve subproblem (69) with the respect to the solution obtained in step 2).
4. If (70) is satisfied, stop the iterative process of the Benders decomposition and output the optimal imputation. Otherwise, generate constraint (71) for master problem (68). Then, repeat steps 2) to 4).

## V. CASE STUDIES

The proposed joint electricity and data trading is applied on the 3-prosumer and the 32-prosumer testing systems, respectively. The numerical tests are performed using a Matlab 2020a platform in Gurobi 9.1 with Intel Core (TM) i5-1135G7 CPU (2.4 GHz) and 16 Gb of memory.

## A. 3-Prosumer System

Figure 4 depicts the 3-prosumer testing system. Prosumers 1 and 2 are equipped with one MT and one DRG, while prosumer 3 only has one DRG. The total load and the total forecasted generations of all DRGs are presented in Figure 5. The capacity of MT in prosumer 1 is larger than that of prosumer 2. All prosumers can trade energy and reserve with the market operator. The trading price with the market operator is shown in Figure 6, which is obtained from the utility company in the United States [28]. The other parameters like generation cost coefficients can be found in Ref. [7].

Two cases are provided to analyze the proposed joint electricity and data trading.

**Case I**: Prosumer only trade electricity (i.e., energy and reserve) with market operator.

**Case II**: Prosumer only trade electricity with their counterparts, but they do not trade or share data.

**Case III**: The proposed joint electricity and data trading, in which prosumers can trade electricity and data with each other.

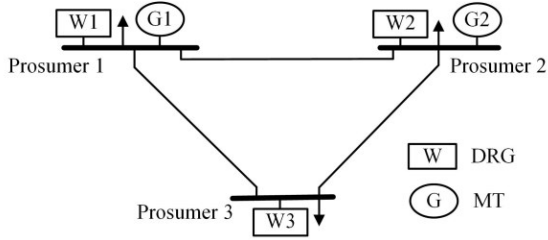

Fig. 4. 3-Prosumer system

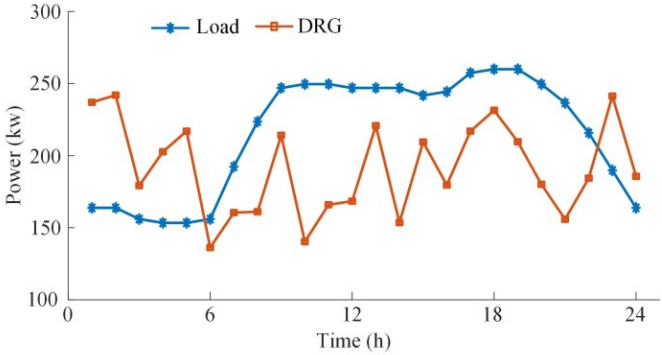

Fig. 5. Total load and total output of DRGs

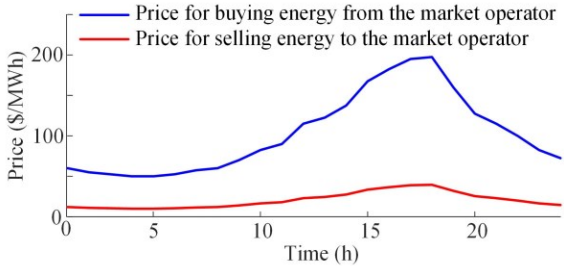

Fig. 6. Price for trading energy with the market operator

The pay allocation of these three cases is provided in Table I. Prosumers can only purchase electricity at a high price from the market operator and sell electricity to the market operator at a low price, which leads to a lower payoff. In Case II, prosumers prefer to trade electricity with their peers rather than the market operator since the electricity provided by their peers is cheaper than that of the market operator. Accordingly, traded electricity with the market operator is decreased while the total payoff is increased in this case. In Case III, prosumers can trade data for reducing the forecast error. More accurate forecast reduces the need on the reserve, and prosumers thus do not purchase reserve from the market operator in Case III. Additionally, prosumers do not need to curtail their loads for providing reserve since more accurate forecast is introduced, which increases their electricity demands. Therefore, prosumers require more electricity from the market operator. In summary, data trading enables more flexible operation by reducing the need on reserve, thus leading to the highest total payoff among these three cases.

Each prosumer's payoff in Cases I-III is depicted in Figure 7. Compared with Case I, prosumers 1 and 2 can provide energy and reserve to Prosumer 3 to accommodate uncertain DRG 3's generation using their MTs in Case II. Therefore, the payoffs of prosumers 1 and 2 are increased. Prosumer 3's payoff is also increased since it can obtain cheap electricity from its peers. In Case III, all prosumers share data for enhancing the forecast and reduce the requirement on reserve. Therefore, all prosumer's payoffs are increased after employing electricity and data trading, indicating that the validity of the proposed joint trading in incentivizing prosumers to share data.

TABLE I PAYOFF ALLOCATION IN DIFFERENT CASES

| Cases | Total Payoff ($) | Cost for trading energy with market operator ($) | | Cost for buying reserve from market operator($) | |
|---|---|---|---|---|---|
| | | Buying | Selling | Up | Down |
| I | 5,703.65 | 1,838.20 | 25.38 | 413.62 | 261.03 |
| II | 7,126.35 | 1,362.59 | 0 | 53.61 | 37.24 |
| II | 8500.87 | 1,561.32 | 0 | 0 | 0 |

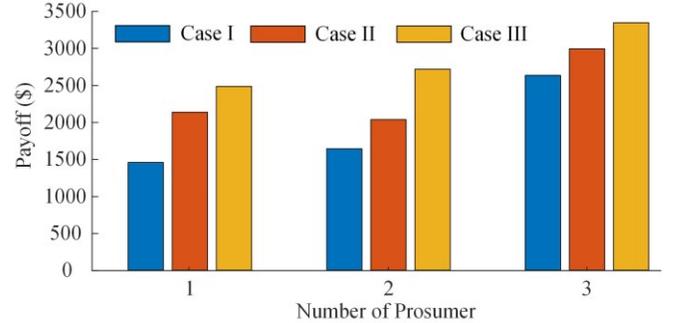

Fig. 7. Payoff of each prosumer in different cases

Table II and Figure 8 compares the proposed imputation method with the most popular imputation methods (i.e., Shapley value and Nucleolus) used in the cooperative game. Table II provides each prosumer's payoff obtained by different imputation methods. Shapley value calculates the imputation of each prosumer based on its expected marginal contribution. However, the imputation obtained by Shapley value may not belong to the *core*. In addition to marginal contribution, Nucleolus takes the fairness and stability of grand coalition into consideration, thus guaranteeing that the imputation belongs to the core. The proposed method can achieve the same solution as the Nucleolus method, indicating the effectiveness of the proposed method. Figure 8 presents each coalition's excess obtained by different imputation methods. According to the definition of excess in (45), each coalition's excess must be negative for incentivizing prosumers to join the grand coalition.

The excesses of coalitions 4 and 5 obtained by the Shapley value are positive. As a result, the members in these coalitions may not be willing to join the grand coalition since they receive less payoff in the grand coalition. The proposed method and Nucleolus ensure the negative excess for each coalition and guarantee the stability of the grand coalition.

Table III illustrates the iterative process of the proposed Benders decomposition for calculating the imputation. The proposed method calculates the imputation through iteration rather than enumeration. The optimal imputation is obtained after 4 iterations by using the proposed method. Compared with the existing imputation methods (i.e., Shapley value and Nuclelous) which require solving at least $2^n-1$ optimization problems, the proposed method mitigates the computation burden by solving less optimization problems. More detailed analysis for the scalability of the proposed method will be provided in the next section.

TABLE II IMPUTATION OBTAINED BY DIFFERENT IMPUTATION METHODS

| Methods | Prosumer's payoff ($) | | |
|---|---|---|---|
| | 1 | 2 | 3 |
| Shapley value | 2,261.73 | 2,621.26 | 3,617.90 |
| Nucleolus | 2,471.65 | 2,702.19 | 3,327.03 |
| Proposed method | 2,471.65 | 2,702.19 | 3,327.03 |

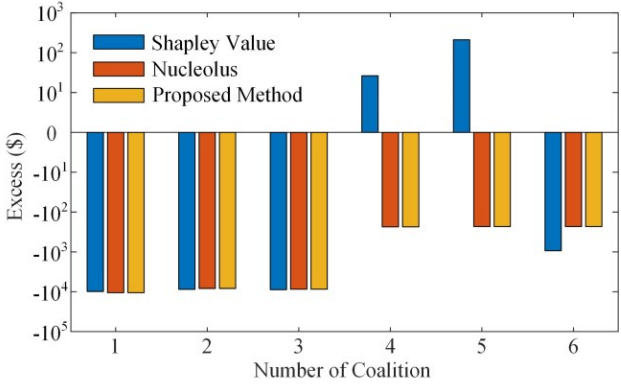

Fig. 8. Excess obtained by different imputation methods

TABLE III ITERATIVE PROCESS OF THE PROPOSED SOLUTION

| Iterations | Prosumer's payoff ($) | | |
|---|---|---|---|
| | Prosumer 1 | Prosumer 2 | Prosumer 3 |
| 1 | 0 | 0 | 8,500.87 |
| 2 | 4,371.86 | 1,635.72 | 2,493.28 |
| 3 | 2,494.07 | 2,574.61 | 3,432.18 |
| 4 | 2,471.65 | 2,702.19 | 3,327.03 |

To further illustrate how prosumers are incentivized to share data using the proposed imputation method, two cases are provided as follow:
- ET: Prosumers share electricity and data. However, the total payoff is imputed to each prosumer based on its contribution in electricity only.
- EDT: The proposed imputation method which imputes the total payoff to each prosumer based on the contribution in both electricity and data.

In the proposed 3-prosumer test system, prosumer 2 contributes most to reducing the uncertainty set while prosumer 3 contributes the least. The total payoffs in these two cases are the same (i.e., 8500.87) since all prosumers trade electricity and data to maximize the total payoff. However, each prosumer's payoffs shown in Figure 9 in these two cases are different. In ET, the payoff is calculated based on the contribution in electricity while the contribution in data is ignored.

Accordingly, prosumers 1 and 3 receives more payoff in this case compared with that in the EDT, while prosumer 2 receives less payoff. Such imputation method is not fair to prosumer 2 since its data contribution in reducing the uncertainty set is ignored. In consequence, prosumer 2 may be reluctant to share its data or choose to share less accurate data. The proposed imputation method incentivizes participants to trade data by accurately recognizing the data value and providing enough payoff for the data, thus ensuring the market sustainability.

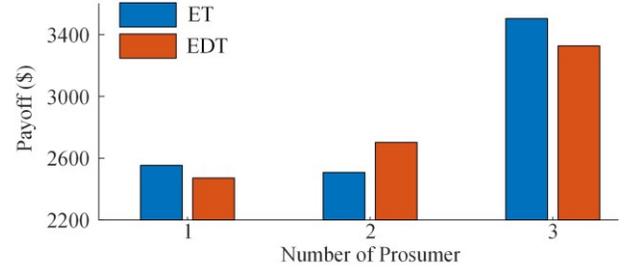

Fig. 9. Payoff of each prosumer in different cases

### B. 32-Prosumer Testing System

A 32-prosumer system is used to test the proposed model. Table IV provides the computational performance of the proposed method. The number of coalitions is exponential growth as the number of prosumers increases. When the number of prosumers reaches 32, there are $4.29*10^9$ coalitions. Conventional imputation methods like Shapley value and Nucleolus are impossible to be applied in practice since they need to enumerate enormous coalitions. The proposed imputation method significantly alleviates the computation burden by avoiding the enumeration. When the number of prosumers reaches 32, only 47 iterations are required to obtain the imputation.

TABLE IV. COMPUTATIONAL PERFORMANCE FOR DIFFEREENT NUMBER OF PROSUMERS

| Number of prosumers | Numbers of coalitions | Iterations | Solution time (s) | |
|---|---|---|---|---|
| | | | Master problem | Subproblem |
| 8 | 256 | 10 | 6.4 | 245.8 |
| 16 | 65,536 | 17 | 10.1 | 691.7 |
| 24 | $1.68*10^7$ | 32 | 18.7 | 2,831.6 |
| 32 | $4.29*10^9$ | 47 | 25.3 | 4,853.2 |

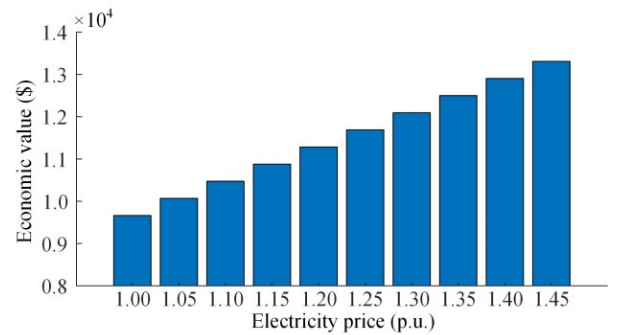

Fig. 10. Economic value of data trading as a funtion of electricity price

Figure 10 depicts the extra economic value created by data trading (i.e., the difference between the total payoff of Case III and that of Case II) as a function of the electricity price. It can be observed that the economic value created by introducing data trading increases as the electricity price increases. This is because data trading improves the forecast accuracy and reduces requirement for importing electricity, thus creating more economic value when electricity price is high. It can also be observed that the economic value of data and electricity are interdependent. Therefore, the joint trading is required to accurately price the data.

## VI. Conclusion

In this paper, a scalable cooperative game for joint electricity and data trading among multiple prosumers is proposed. The market operator first collects relevant parameters from all prosumers to reduce the forecast error of DRGs. Then, the market operator maximizes the grand coalition's payoff considering uncertain generation of DRGs and further assigns the total payoff to all participating prosumers. The grand coalition's mathematical model is formulated as a SOCP problem by using the affine robust approach. The stability of the grand coalition is mathematically proved. A two-stage nonlinear and nonconvex imputation problem is proposed to impute the total payoff to each prosumer, which is further converted into a MISOCP problem and solved by the Benders decomposition. Numerical results on a 3-prosumer and 32-prosumer systems illustrate the validity of the proposed joint electricity and data trading in incentivizing prosumers to cooperate for trading electricity and data, and enhances the scalability of the imputation in the cooperative game. Our future work will study the blockchain to enable data circulation.